\batchmode
\documentstyle[preprint,revtex]{aps}

\begin{document}
\draft
\begin{title}
Static Charged Black Holes in $(2+1)$-Dimensional Dilaton Gravity
\end{title}
\author{K.C.K. Chan and R.B. Mann}
\begin{instit}
	Department of Physics
	University of Waterloo
	Waterloo, Ontario
	CANADA N2L 3G1
\end{instit}
\begin{center}April 21, 1994\end{center}
\begin{center}{WATPHYS TH-94/01}\end{center}

\begin{abstract}
A one parameter family of static charged black hole solutions in
$(2+1)$-dimensional general relativity minimally coupled to a dilaton
$\phi\propto ln({r\over\beta})$ with a potential
term $e^{b\phi}\Lambda$ is obtained.
Their causal strutures are investigated, and
thermodynamical temperature and entropy are computed. One
particular black hole in the family has the same thermodynamical
properties as the Schwarzschild black hole in $3+1$ dimensions.
Solutions with cosmological horizons are also discussed. Finally,
a class of black holes arising from the dilaton with
a negative kinetic term (tachyon) is briefly discussed.

\end{abstract}

\section{Introduction}

$(2+1)$ dimensional gravity continues to be a source of fascination
for theorists, primarily because of the potential insights
into quantum gravity that it offers. The lower-dimensional setting affords
a significant amount of technical simplification of the gravitational field
equations, bringing into sharper focus conceptual issues that are often
obscured in the more complicated $(3+1)$-dimensional case. However such
technical simplicity is not without consequence: for example, in the case
of $(2+1)$-dimensional general relativity the metric outside of a matter
source of finite spatial size is locally flat, and the mass affects the
space-time only globally, seemingly implying that $2+1$ gravity will not
admit black hole solutions.

Fortunately the actual situation is significantly more interesting.
Recently, Banados {\sl et al} showed that by identifying certain points of
$(2+1)$-dimensional anti-de Sitter space, one obtains a solution with
almost all usual features of a black hole [1]. This ``$BTZ$'' black hole
has so far attracted much interest in its classical, thermodynamical
and quantum properties [2]. In particular, it is also a
solution to a low energy string theory [3]. Furthermore, by taking the
product of the $(1+1)$-dimensional string-theoretic black hole  of Mandal,
Sengupta and Wadia [4] with ${\bf{S^1}}$ another $(2+1)$-dimensional black
hole solution to the string theory may be obtained (hereafter referred
as $2+1$ $MSW$ black hole) -- if a product with ${\bf {R}}$ is taken instead,
one gets a black string [5]. One can further conformally transform [5,6] this
black hole such that the transformed metric is a solution to an
Einstein-Maxwell-dilaton action with special values of the
couplings. Since causal structures are preserved under conformal
transformations, as long as the transformations are at least finite outside
(and on) the event horizon, one may obtain in this manner another black
hole solution to the non-vacuum Einstein field equations in $(2+1)$
dimensions.

Motivated by the above, we consider in this paper black hole solutions to
the following Einstein-Maxwell-dilaton action
$$
S=\int{d^{3}x\sqrt{-g}(R - {B\over 2}(\nabla\phi)^2
  - e^{-4a\phi}F_{\mu\nu}F^{\mu\nu} + 2e^{b\phi}\Lambda)}, \eqno(1)
$$
with arbitrary couplings $\Lambda$, $a$, $b$ and $B$, where $R$ is the
Ricci scalar, $\phi$ is the dilaton field and $F_{\mu\nu}$ is the usual
Maxwell field. We find that in addition to the aforementioned black holes
in $2+1$ dimensions, there exists a one parameter family of static non-
asymptotically anti-de Sitter charged black hole solutions
, as well as solutions which admit cosmological horizons.
We shall still refer to $\Lambda$ as the
cosmological constant, although in the presence of a non-trivial dilaton,
the space does not behave as either de Sitter ($\Lambda<0$) or anti-de
Sitter space ($\Lambda>0$). The constants $a$ and $b$ govern the coupling
of $\phi$ to $F_{\mu\nu}$ and $\Lambda$ respectively. Without loss of
generality, the parameter $B$ in (1) can be set to 8 as usual. However
we shall leave $B$ as an arbitrary parameter so as to more conveniently
permit its continuation to negative values. The $BTZ$ and $MSW$ black holes
are two of the solutions of the field equations which follow from
(1), obtained for particular values of the
couplings $a$ and $b$. Of course eq.(1) can be viewed as general
relativity with an unusual matter Lagrangian. For example, one can
easily see that the local energy density in the perfect fluid form is
negative when $\phi=constant$ and $\Lambda>0$.

The corresponding action for the low energy string theory
can be obtained by setting $B=8$, $b=4$, $a=1$
and carrying out the conformal transformation [6]
$$
g^{S}_{\mu\nu}=e^{{4\phi\over(n-1)}}g^{E}_{\mu\nu}, \eqno(2)
$$
where $S$ and $E$ denote the string and Einstein metrics respectively, with
$n$ the number of spatial dimensions. The corresponding string action is
$$
S=\int d^{3}x\sqrt{-g^S}e^{-2\phi}(R[g^S]
+4(\nabla\phi)^2-F^2+2\Lambda), \eqno(3)
$$

A number of other special cases of the action (1) have previously been
considered by a number of authors. Barrow {\sl et al} obtained the most
general static and circularly symmetric solutions for $\Lambda=0$,
as well as a solution with a non-vanishing $\Lambda$,
$B=2$, $F_{\mu\nu}=0$, $b=-\sqrt{2}$ and $a=0$ [7].
They did not obtain any black hole
solution. Burd and Barrow [8],
as well as Muslomov [9], discussed action
(1) in the context of pure scalar field cosmology; several exact solutions
were obtained for a $(2+1)$-dimensional spatially flat $FRW$ model with
$B=2$ and $b<0$. Shiraishi derived static multi-centered solutions for
the $\Lambda=0$ case [10]. We will later show that his solutions actually
correspond to those of massless charged particles. Note that Maki and
Shiraishi considered a similar type of action in more than two spatial
dimensions [6]. They found exact solutions for a configuration of multiple
black holes. It is also worthwhile to point out that the action (1)
describes a Liouville-type gravity [11]; in the absence of
electromagnetism, the action is an extension of the usual Liouville action
in curved spacetime to $(2+1)$ dimensions. Exact black hole solutions to
the lowest dimensional ({\sl i.e.} $1+1$) Liouville-type gravity have
recently been found [11].

The organization of this paper is as follows. In section $2$ we adopt a
static and circularly symmetric ansatz and then write down and solve the
field equations. In section $3$, we consider the solutions with $B>0$ (this
corresponds to positive kinetic energy in flat space for the dilaton). The
quasilocal mass [12] associated with these solutions is computed and a mass
parameter identified. For positive quasilocal mass, we obtain a
one-parameter family of static charged black hole solutions. We discuss
their causal structures and in section $4$ compute their relevent
thermodynamic properties ({\sl i.e.} Hawking temperature $T_H$ and entropy
$S$). We will see how a non-trivial dilaton alters
the causal structures and thermodynamical properties with respect to
the BTZ black hole. One black hole solution has thermodynamical
properties which are the same as the $(3+1)$-dimensional Schwarzschild black
hole; in
both cases, $T_H\propto (mass)^{-1}$. In section $5$, we consider
solutions with cosmological horizons, and in section $6$, we briefly
investigate the case $B<0$. In this latter case we find a solution
corresponding to a
massless charged black hole. We find no black hole solutions of positive
quasilocal mass when $\Lambda\le 0$. We summarize our work in a concluding
section.

Our conventions are as in Wald [13]; we set the gravitational
coupling constant $G$, which has a dimension of an inverse
energy, equal to unity.

\vskip 0.5 true cm

\section{Exact Solutions}

Varying (1) with respect to
the metric, Maxwell and dilaton fields respectively yield (after
some manipulation)
$$
R_{\mu\nu}={B\over 2}\nabla_{\mu}\phi\nabla_{\nu}\phi
	   + e^{-4a\phi}(-g_{\mu\nu}F^2 + 2F_{\mu}{}^\alpha F_{\nu\alpha})
	   - 2g_{\mu\nu}e^{b\phi}\Lambda, \eqno(4)
$$
$$
\nabla^{\mu}(e^{-4a\phi}F_{\mu\nu})=0, \eqno(5)
$$
$$
{B\over 2}(\nabla^{\mu}\nabla_{\mu}\phi) + 2ae^{-4a\phi}F^2
 + be^{b\phi}\Lambda=0. \eqno(6)
$$
We wish to find static, circularly symmetric solutions to these equations
that have a regular horizon. In $2+1$ dimensions,
the most general such metric has
two degrees of freedom [14], and can be written in the form
$$
ds^2=-U(r)dt^2+{dr^2\over U(r)}+H^2(r)d\theta^2. \eqno(7)
$$
This is different from the usual ansatz for a circularly
symmetric metric, but turns out to simplify the calculations.
For an electric point charge
$F_{\mu\nu}$ has just one independent component,
$F_{01}=-F_{10}=e^{4a\phi}f(r)$ (the
magnetic field is a scalar in $2+1$ dimensions and will
not be considered here). Adopting the metric (7), eq.(5) implies that
$$
f(r)={q\over H(r)}, \eqno(8)
$$
where $q$ is an integration constant. Now, eqs.(4)-(8)
together yield
$$
U'' + U'{H'\over H}=4e^{b\phi}\Lambda,
\eqno(9)
$$
$$
{H''\over H}=-{B\over 2}(\phi')^2, \eqno(10)
$$
$$
U'{H'\over H} + U{H''\over H} = 2\left(e^{b\phi}\Lambda-{e^{4a\phi}q^2\over
H^2}\right),
\eqno(11)
$$
$$
{B\over 2}[U(\phi'{H'\over H}+\phi'') + U'\phi'] - {4ae^{4a\phi}q^2\over H^2}
+ be^{b\phi}\Lambda=0, \eqno(12)
$$
where $'={d\over dr}$, denoting the ordinary derivative with respect to $r$.
The first one is the $R_{tt}$ equation. $R_{tt}$ and $R_{rr}$ together
yield eq.(10) and $R_{\theta\theta}$ yields (11).
The last equation is the local energy conservation
equation, $\nabla^{\mu}T_{\mu\nu}=0$. From eq.(10) it is easy to see
that $H^2\propto r^2\iff\phi=constant$. Thus one generally
cannot have $g_{tt}={-1\over g_{rr}}$ and $g_{\theta\theta}=r^2$
simultaneously in $2+1$ dimensions when one has a non-trivial solution for
the dilaton.

For the $BTZ$ charged black hole, one has $U(r)=(-m+{\Lambda}r^2-
2Q^2ln({r\over r_o}))$ [15], $H(r)=r$, and $\phi=a=b=0$. It is easy to
check that the field equations are satisfied. For the $2+1$ $MSW$ black
hole, one can verify that conformal transformation (2) (with $\phi=-{1\over
4}ln({r\over \beta})$, $\beta=constant$) yields the following Einstein metric
$$
ds^2=-(8\Lambda{\beta}r-2m\sqrt{r})dt^2
+ {dr^2\over (8\Lambda{\beta}r-2m\sqrt{r})}
+ {\gamma^2}rd\theta^2, \eqno(13)
$$
where $m$ is the square
root of the mass per unit length and $\gamma$ is an integration constant
with dimension $[L]^{1\over 2}$. Eq.(13) is an exact
solution to eqs.(9)--(12) with $q=0$, $b=4$, and $B=8$.
If we perform the duality
transformation to the corresponding string metric, a charged
solution can be obtained [5]. In terms of Einstein metric, we have
$$
ds^2=-(8\Lambda{\beta}r-2m\sqrt{r}+8Q^2)dt^2 + {dr^2\over
(8\Lambda{\beta}r-2m\sqrt{r}+8Q^2)}+\gamma^2rd\theta^2, \eqno(14)
$$
where $Q$ is the charge. It can be checked that metric (14) is
also an exact solution to eqs.(9)--(12) when $Q^2={q^2\beta\over\gamma^2}$,
$b=4a=4$ and $B=8$.

We see that when $H^2=\gamma^2r$ and $H^2=r^2$ in the Einstein metric (7),
one gets the $MSW$ black hole (13) (or (14)) and the $BTZ$ black hole
respectively. Consider, then, the ansatz
$$
H^2=\gamma^2r^N. \eqno(15)
$$
Since $r$ and $H^2$ have a dimension of $[L]$ and $[L]^2$ respectively,
the dimension of $\gamma$ is $[L]^{{2-N\over 2}}$.
In addition, we further assume
$$
\phi=k ln{\left({r\over\beta}\right)}, \eqno(16)
$$
where $k$ is a real number.
Inserting (15) and (16) in (9)--(12) we get the exact solutions
$$
ds^2=-U(r)dt^2+{dr^2\over U(r)} + \gamma^2 r^Nd\theta^2,
\eqno (17a)
$$
with
$$
U(r) = \left[Ar^{1-{N\over 2}} + {8\Lambda\beta^{2-N}\over (3N-2)N}r^N
+ {8Q^2\over N(2-N)}\right] \eqno(17b)
$$
and where
$$
k=\pm\sqrt{{N(2-N)\over 2B}}, \eqno(18)
$$
$$
4ak=bk=N-2,\qquad 4a=b. \eqno(19)
$$
The constant of integration $q$ in eq.(8) is related to the
charge $Q$ via $Q^2={q^2\beta^{2-N}\over\gamma^2}$, whereas
$A$ is (as we shall subsequently demonstrate) a constant of integration
proportional to the quasilocal mass.

The solutions (17) depend on the dimensionless couplings $a$ and $b$
(or alternatively $N$) and on the integration
constants $A$ and $\gamma$. By performing the coordinate transformation
${\gamma}^2r^N\rightarrow r^2$, (17) becomes
$$
ds^2=-\left(Ar^{{2\over N}-1}
+ {8{\Lambda}r^2\over (3N-2)N}+{8Q^2\over (2-N)N}\right)dt^2
$$
$$
+ {4r^{{4\over N}-2}dr^2\over {N^2{\gamma}^{4\over N}\left(Ar^{{2\over N}-1}+
{8{\Lambda}r^2\over (3N-2)N} + {8Q^2\over (2-N)N}\right)}}+r^2d\theta^2.
\eqno(20a)
$$
where from now on $r$ denotes the usual radial co-ordinate. We have
absorbed $\gamma^{1-{2\over N}}$ into the constant $A$
and normalized $\beta^{2-N}\gamma^{-2}$ to $1$. Now, $Q^2=q^2$ and
$$
\phi={2k\over N}ln\left({r\over\beta(\gamma)}\right).
\eqno(20b)
$$

Before proceeding further we note the following.
First, it is obvious that as $r\rightarrow\infty$,
$\phi\rightarrow\infty$ too. However, the kinetic term,
Maxwell term, and the potential term in action (1)
all vanish in that limit when $2\geq N>0$ ({\sl i.e.}, $B>0$).
On the other hand, when $B<0$, they all diverge. Second, for
non-vanishing $Q$, when $N=2$ (or equivalently, $a=b=0$)
the metric diverges and the
present solution has no smooth connection to the $N=2$ case, similar to the
situation in ref.[10]. (This can easily be seen as follows: one can
write the charged metric coefficient $U$ as $U_{Q=0}+h(r)$; for $N=2$,
eqs.(9)--(12) imply $h(r)=-2Q^2ln({r\over r_o})$. When $N\ne 2$, the same
set of equations forces $h(r)$ to be a constant, inversely proportional to
$(2-N)N$. A smooth connection to $N=2$ case is possible only when $Q=0$.)
Third, when $N=1$, (20a) reduces to the $2+1$ $MSW$ charged black hole.
Fourth, if $N=2$, $Q=0$ and $\Lambda=0$, one obtains the usual vacuum one
particle solution in $2+1$ dimensions [16]. Fifth, if both $\Lambda$ and
$A$ vanish, (20a) reduces to the Shiriashi single particle solution [10] in
Schwarzschild form. As we will later show that $A$ is linearly proportional
to the quasilocal mass, we see that Shiriashi's solution in fact describes
a static massless charged particle. Sixth, when $Q=A=0$ (the vacuum
solution), the metric (20a) does not have an infinitely long throat for small
$r$ except $N=2$ ($BTZ$ case). Also, (20a) fails to fullfill the falloff
conditions given in [17] for asymptotically anti-de Sitter spaces. Finally,
as mentioned previously, the action (1) is related to the string action (3)
by a conformal transformation (2) if $b=4a=4$ and $B=8$. From (18) and (19)
we see that this choice of parameters forces $N=1$. In addition,
we have mentioned earlier that for the uncharged $N=2$ (BTZ) black hole
it is a solution to string theory
(with an addition of a three form $H_{{\rho}{\mu}{\nu}}$,
see ref.[3] for details.). If $N$ differs from $1$ or $2$,
the solution (20) ``decouples'' from string theory.

\section{Black hole solutions for $2>N>{2\over 3}$, $\Lambda>0$,
		 $A<0$}

In this section we seek to determine under what circumstances the set of
solutions (20a) has black hole event horizons. The location of the
horizon(s) will be given by the real positive roots of $g_{tt}=0$.  However,
there exists no general method to obtain roots for a general value of $N$.
In this section, we will restrict ourselves to the case where both $B$ and
$\Lambda$ are positive; the former condition implies $2>N>0$ ($N\ne{2\over
3}$, see (18) and (20a)). We first investigate the case $2>N>{2\over3}$.
The other case, ${2\over 3}>N>0$ will be shown in section 5 to have
cosmological horizons.

We first determine the quasilocal mass to the solution (20a) with
$2>N>{2\over3}$. For a static metric in $(2+1)$ dimensions of the
form
$$
ds^2=-W^2(r)dt^2+{dr^2\over V^2(r)}+r^2d\theta^2,
\eqno(21)
$$
and when the matter action contains no derivatives of the
metric, the quasilocal energy $E$ and mass $M$ are respectively
given by [12]
$$
E=2(V_o(r)-V(r)), \eqno(22)
$$
and
$$
M=2W(r)(V_o(r)-V(r)). \eqno(23)
$$
Here $V_o(r)$ is an arbitrary function which determines the zero of
energy relative to some background spacetime and $r$ is the radius
of the spacelike hypersurface boundry; we will choose
$V_o(r) = 1/g_{rr}(r;A=0)$ for $g_{rr}$ given by (20a).
For $2>N>{2\over 3}$, eqs.(20a) and (23), yield
$$
M=N{\gamma}^{2\over N}r^{1-{2\over N}}\left(Ar^{{2\over
N}-1}+{8{\Lambda}r^2\over(3N-2)N}+{8Q^2\over(2-N)N}\right)^{1\over 2}
\left[\left({8{\Lambda}r^2\over(3N-2)N}+{8Q^2\over(2-N)N}\right)^{1\over 2}
\right.
$$
$$
-\left.
\left(Ar^{{2\over N}-1}+{8{\Lambda}r^2\over(3N-2)N}
+{8Q^2\over(2-N)N}\right)^{1\over2}\right].
\eqno(24)
$$
Since $2>({2\over N}-1)$ the upper bound of the radius $r$ is infinity
regardless of the sign of $A$. Taking this limit we find
$M=-{AN{\gamma}^{2\over N}\over 2}$. It is obvious that if $M>0$
(positive mass), $A<0$ and vice versa. For convenience, we absorb the
constant $\gamma^{{2\over N}}$ into $M$ (so that $M$ has dimension
$[L]^{{N-2}\over {N}}$).
Thus one can identify
$$
A=-{2M\over N}. \eqno(25)
$$
Thoughout this paper, we will restrict $M\geq 0$ ($A\leq 0$).
The quasilocal energy can be calculated similarly.
As $r\to\infty$, one can check that $E\to 0$ in (20a),
similar to the $BTZ$ black hole [12].

Substituting (25) into metric (20a), the equation $-g_{tt}=0$
gives
$$
-{2M\over N}r^{{2\over N}-1}+{8{\Lambda}r^2\over(3N-2)N}+{8Q^2\over(2-N)N}=0.
\eqno(26)
$$
Note that with our choice of $N$, the middle and
last terms are always positive. The first term is always negative.
Event horizons exist only in the situations $2>N>2/3$ and $\Lambda>0$.

The location(s) of the event horizon(s) in the coordinates (20a) will in
general depend upon $M$, $\Lambda$, $Q$ and $N$. Before proceeding to this
general case we consider some illuminating examples: there exist five
rational numbers within the range $2>N>{2\over 3}$ such closed form solutions
to
(26) may be obtained. They are $N={6\over5}, {6\over7}, {4\over3},
{4\over5}$ and $1$. The first two give a cubic equation to (26), the
next two give a quartic one, and the last a quadratic. This latter case
corresponds to the string-theoretic black hole discussed in the previous
section. We consider the other four in turn. For each of these the metric
is given by eq.(20a), with $A$ as in eq.(25).

Consider first the case $N={6\over 5}$, for which the solutions to (26) are
the zeros of the following equation
$$
y=r^2-{2M\over5\Lambda}r^{{2\over 3}}+{2Q^2\over\Lambda}. \eqno(27)
$$
where we note that both ${2M\over5\Lambda}$ and ${2Q^2\over\Lambda}$ are
positive. A straightforward graphical analysis shows that
there are three possible cases to eq.(27). The first case
corresponds to $y_{min}>0$ which is equivalent to
the condition ${Q^2\over\Lambda}>({2M\over15\Lambda})^{{3\over 2}}$, for
which (27) has no real positive root. There is no black hole: the charge
$Q$ is too large with respect to a given $M$ and $\Lambda$, and the
singularity is naked. The second case corresponds to $y_{min}=0$,
which is equivalent to ${Q^2\over\Lambda}=({2M\over15\Lambda})^{{3\over 2}}$,
and has one real positive root which is located
at $r_{min}=({2M\over15\Lambda})^{{3\over4}}$,
corresponding to an extremal black hole. Finally, if $y_{min}<0$, then
${Q^2\over\Lambda}<({2M\over15\Lambda})^{{3\over 2}}$ there are two real
positive roots and the black hole spacetime has both an outer
and inner horizon. Generally, eq.(27) has at most two real
positive roots. The above situations are qualitatively the same as the
Reissner-Nordstrom black hole in $3+1$ dimensions. In general, the
(positive) roots of (27) are given by
$$
r=\left(-{\sqrt{8M\over\ 15\Lambda}}cos\theta\right)^{{3\over 2}},\qquad
cos(3\theta)=\left({15\Lambda^{{1\over3}}\over2M}\right)^{3\over2}Q^2.
\eqno(28)
$$
which define the location of the horizons.
When $Q=0$ (implying $cos(3\theta)=0$) the horizon is at $r_{h}=
({2M\over5\Lambda})^{{3\over 4}}$; for the extremal case,
$cos(3\theta)=1$ which implies $r_h=({2M\over15\Lambda})^{{3\over 4}}$.
In general the location of the outer horizon is determined by the largest
positive value of $r$ in (28). For example in the special case that
$Q^2\sqrt{\Lambda}
={1\over\sqrt{2}}({2M\over 15})^{{3\over 2}}$ the outer horizon is
$r_{+}=({15Q^2\over M})^{{3\over2}}$ and the inner one is
$r_{-}=[({15Q^2\over 2M})({\sqrt3}-1)]^{{3\over2}}$.

We consider next the causal structure of the $N={6\over 5}$ metric.
Since eq.(27) is cubic in $r^{{2\over 3}}$ and has
at most two real positive roots we
proceed by first obtaining the causal structure in the uncharged case, and
then deducing the structure for the $Q\neq 0$ spacetime (a method similar
to ref.[18]). The conformal radial co-ordinate
$r_{*}\equiv\int{\sqrt{-{g_{rr}\over g_{tt}}}}dr$ for $N={6\over 5}$
and $Q=0$ is given by
$$
r_{*}={3\over10\Lambda\alpha\gamma^{5\over 3}}\left[ln\left({r^{{1\over3}} -
\alpha\over r^{{1\over3}}+\alpha}\right)
+2tan^{-1}\left({r^{{1\over3}}\over\alpha}\right)\right],
\eqno(29)
$$
where $\alpha=({2M\over 5\Lambda})^{{1\over4}}=r^{{1\over 3}}_{h}$.
The advanced and retarded null co-ordinates are defined as
$u=t-r_{*}$, $v=t+r_{*}$, as usual. Defining the Kruskal co-ordinates as
$$
{\cal U}=-{3\over 5\Lambda\alpha\gamma^{5\over 3}}
e^{-{5\Lambda\alpha\gamma^{5\over 3}\over 3}u},\qquad
{\cal V}={3\over 5\Lambda\alpha\gamma^{5\over3}}
e^{{5\Lambda\alpha\gamma^{5\over 3}\over 3}v} \eqno(30)
$$
the metric can be written as
$$
ds^2=-\left({5Mr^2\over 3}\right)
\left({1\over \alpha^4}-{1\over r^{4\over 3}}\right)
\left({r^{1\over 3}+\alpha\over r^{1\over 3}-\alpha}\right)
e^{-2tan^{-1}\left({r^{1\over 3}\over\alpha}\right)}d{\cal U}d{\cal V},
\eqno(31)
$$
and
$$
{\cal U}{\cal V}=-\left({3\over
5\Lambda\alpha\gamma^{5\over 3}}\right)^2 \left({r^{{1\over3}}-\alpha\over
r^{{1\over3}}+\alpha}\right) e^{2tan^{-1}({{r^{1\over3}\over\alpha}})}.
\eqno(32)
$$
The Penrose diagram is given in fig.(1). As $r\to\infty$,
${\cal U}{\cal V}\to -({3\over 5\Lambda\alpha\gamma^{5\over 3}})^2 e^{\pi}$,
which corresponds to a vertical ({\sl i.e.} a timelike) line in the Penrose
diagram. The horizon is at $r=r_{h}=\alpha^3$ which implies ${\cal U}{\cal
V}=0$. On passing from $r>r_{h}$ to $r<r_{h}$, the metric (31) changes
sign and one has to transform ${\cal U}\to{\tilde{\cal U}} =-{\cal U}$ so
that the time direction is still vertical. Now, ${\tilde{\cal U}}{\cal
V}=-({3\over 5\Lambda\alpha\gamma^{5\over 3}})^2$ when $r=0$, indicating
that the singularity is timelike as well. It is lenghtly
but straightforward to check that the Ricci and Kretschmann
scalars diverge at $r=0$.
Thus the spacetime has a timelike scalar
curvature singularity. (A similar causal structure has been found for
a class of $(1+1)$ dimensional black holes in ref.[19]). A test
particle may travel along a future directed timelike curve from region $I$,
passing through $r_h$ to region $II$. Without hitting the singularity, it
can re-emerge from region $II$ and enter another region $I$. In fact, one
has an infinitum of regions $I$ and $II$. When charge is added, the manifold
splits into three different regions: region $I$ ($r>r_{+}$), region $II$
($r_{+}>r>r_{-}$) and region $III$ ($r_{-}>r$). In this case, whenever
one crosses the horizons
$r_{+}$ or $r_{-}$, the space and time co-ordinates interchange roles
and the singularity is spacelike. We can deduce that the charged
black hole has causal structure looks like fig.(2), and for the extremal
case, it becomes fig.(3). These causal structures are similar to some
of those found in ref.[20], where the authors investigated dimensionally
continued Lovelock black holes with a cosmological
constant $\Lambda >0$ in both odd and even
spacetime dimensions. It is interesting to note that in their
static charged black hole solutions in both odd and even dimensions,
they always have a term
${\Lambda}r^2$, similar to ours (see eq.(20a)). It is tempting to consider
the same kind of dimensional continuation with the presence of a dilaton;
however we will not discuss it here.

We next turn to our attention to the case $N={6\over 7}$. The analogue of (27)
is now
$$
y=r^2-{M\over 7\Lambda}r^{{4\over 3}}+{Q^2\over 2\Lambda}.
\eqno(33)
$$
As with the previous case, a graphical analysis indicates that
there are three kinds of curves, corresponding to
${Q^2\over \Lambda}>({2M\over 21\Lambda})^3$ (a naked singularity),
${Q^2\over \Lambda}=({2M\over 21\Lambda})^3$ (an extremal black hole)
and ${Q^2\over \Lambda}<({2M\over 21\Lambda})^3$ (a double-horizon
black hole) respectively. The roots are implicitly given by
$$
r=\left[{M\over 21\Lambda}(1-2cos\theta)\right]^{{3\over 2}},\qquad
cos(3\theta)=-1+\left({21\over M}\right)^3\left({Q\Lambda\over 2}\right)^2.
\eqno(34)
$$
It is easy to see that $Q=0$ gives $r_{h}=({M\over 7\Lambda})^{{3\over2}}$.
For the extremal case, one has $cos(3\theta)=1$ which yields
a positive $r_{h}=({2M\over 21\Lambda})^{{3\over2}}$. As an intermediate
example, if one sets $Q\Lambda=2({M\over 21})^{3/2}$, then the outer and inner
horizons are $r_{+}= ({21Q^2\over 4M})^{3\over4}(\sqrt3+1)^{{3\over2}}$
and $r_{-}= ({21Q^2\over 4M})^{3\over4}$.

Following the co-ordinate transformations outlined
above, we can obtain the causal structures. We only show
the uncharged case (fig.(4)). It is similar to the one for
an extremal Reissner-Nordstrom (RN) black hole in $3+1$ dimensions.
There is a timelike singularity at which the
Ricci and Kretschmann scalars diverge.
The causal structure for generic charged
black holes is obtained from the one for generic
RN black holes by a ${\pi\over 2}$ rotation. For the
extremal one, it is simply obtained through a ${\pi\over 2}$ rotation
of fig.(4). Note that those causal structures are similar to the
ones in ref.[20] for asymptotically flat charged Lovelock
black holes in even dimensions.

Substituting $N={4\over 3}$, ${4\over 5}$ into (26) yields two quartic
equations. The analysis of the roots is similar to the previous two cases,
and we find that there are either two positive roots, one (the extremal
case) or none. The $N={4\over 3}$ black hole,
which has the same causal structures as the $N={6\over 5}$ one,
has the extremal condition
${Q^2\over\Lambda}=({M\over 8\Lambda})^{{4\over3}}$;
the horizon is at $r_{h}=({M\over 8\Lambda})^{{2\over 3}}$.
For the $N={4\over 5}$ one, which has the same causal structures
as the $N={6\over 7}$ black hole,
the extremal condition is ${Q^2\over\Lambda}=({3M\over 40\Lambda})^4$, and
the horizon is at
$r_{h} = ({3M\over 40\Lambda})^2$ respectively.

For general $N$ such that $2>N>{2\over 3}$, a similar graphical
analysis shows that we either get a black hole (generic or extremal)
or a naked singularity. Rewriting (26) as
$$
x^{2\lambda} - \lambda x^2 + \delta(\lambda - 1) = 0, \eqno(35a)
$$
where $\lambda = {2-N \over 2N}$, $x={r\over r_m}$,
$\delta = {Q^2 \over \Lambda r^{2}_m}$ and
$$
r_m=\left[{(3N-2)(2-N)M\over 8N\Lambda}\right]^{{N\over 3N-2}} \qquad .
\eqno(35b)
$$
The extremal condition is when $\delta = 1$, for which $r_h=r_m$ ({\it
i.e.} $x=1$).  For a given $M$ and $\Lambda$ two horizons exist if
$\delta<1$, they coalesce if $\delta = 1$ and if $\delta >1$ there is only a
naked curvature singularity.

So far we have been concerning $\Lambda>0$. When $\Lambda$ flips sign
or vanishes, one can see from metric (20a)
that there is no event horizon for a positive
mass. We also comment that higher-order polynomials (e.g. cubic
or quadric) in the variable in equation $g_{tt}=0$ for
a generic static black hole metric are not uncommon.
For example, in the $(3+1)$-dimensional charged
Schwarzschild de Sitter metric, one has a quartic equation (see, {\sl e.g.}
[21]) whereas in $(1+1)$-dimensional string theory, higher-loop corrections
may lead to higher-order polynomials in $g_{tt}=0$ for a charged black hole
[18] and multiple real and positive roots may be expected. (In (14)
$g_{tt}=0$ is a quadratic equation in $r$ since it corresponds to the
lowest loop-order string theoretic action (3)).

\vskip .5 true cm

\section{Thermodynamics}

An important thermodynamical quantity in a static black hole
is the Hawking temperature $T_{H}$. Given a static and
circularly symmetric black hole metric, $T_{H}$ is given by
$$
T_{H}={|g_{tt}'|\over 4\pi}\sqrt{-g^{tt}g^{rr}} \vert_{r=r_+}.
\eqno(36a)
$$
along with a blueshift factor which may be computed as in ref.[12].
Using eqs.(20a), (25) and (35) the Hawking temperature becomes
$$
T_H={\lambda M\gamma^{2\over N}\over 2\pi r_+}
\left[ \left({r^2_+ \over r^2_m}\right)^{1-\lambda} - 1 \right]
\eqno(36b)
$$
with $r_+$ the location of the (outer) event horizon.
It is easy to check that for the uncharged $BTZ$ black
hole, $T_H={1\over 2\pi}\sqrt{M\Lambda}$ as expected; for the uncharged
$MSW$ black hole, $T_{H}={\gamma^2\Lambda\over \pi}$, independent of mass.

Given the outer horizon $r_+$ as a function of $M$,
$Q$ and $\Lambda$, one can use eq.(36) to get the
Hawking Temperature in terms of these quantitites. For
general $N$ (including the four
cases discussed above) $r_+$ is defined in terms
of these quantities via (35): for a given $\lambda$ and $\delta$,
one can numerically solve (35), obtaining $r_+ = x_0(\lambda,\delta)r_m$,
where $x_0$ is the largest positive root of (35), yielding
$$
T_H={\lambda M\gamma^{2\over N}\over 2\pi x_0(\lambda,\delta) r_m }
\left[ (x^2_0(\lambda,\delta))^{1-\lambda} - 1 \right].
\eqno(36c)
$$
For extremal black holes, $x_0=1$
and $T_{H}=0$. Thus they are stable endpoints of Hawking evaporation.

For $\delta=0$ ({\it i.e.} $Q=0$) (36c) becomes
$$
T_{H}={\gamma^{2\over N}\Lambda\over \pi N}
\left({(3N-2)M\over{4\Lambda}}\right)^{2(N-1)\over 3N-2}  \qquad .
\eqno(37)
$$
Note that (1) describes Einstein-Hilbert action coupled to matter
whose kinetic energy is quadratic; thus for those ``dirty
black holes'' the entropy is still given by $S=4\pi r_{h}$ [22]:
$$
S=4\pi r_{h} = 4\pi \left({(3N-2)M\over{4\Lambda}}\right)^{N\over 3N-2}
\qquad .
\eqno(38)
$$
Using this relation between $S$ and $r_h$, it is easy to check that eq.(37)
can also be derived from the quasilocal energy $E$ given by eq.(22), except
that the blue shift factor ${1\over g_{tt}(r)}$ will be introduced in
eq.(37). From eq.(38), one can also see that $S$ can also be obtained through
the equation ${\partial S \over\partial M} = {1\over T}$ of thermodynamics
of event horizons in $2+1$ dimensions [23] and eq.(37).

When $Q\ne 0$, the relation $S=4\pi r_{+}$ still holds since the addition
of a point electric charge in the matter action in (1) will not change the
fact that the entropy is proportional to the area of the horizon [22].
Given such a relationship between $S$ and $r_+$,
following the argument in ref.[23], one can loosely see
that in $2+1$ dimensions, the area of an event horizon
(generated by null geodesics) never decreases in time if
$R_{\nu\mu}k^{\nu}k^{\mu}\geq 0$, where $k^{\mu}$ is the tangential
vector field of the null geodesics in a congruence. Similar to the
Reissner-Nordstorm and Schwarschild-de Sitter types spacetimes in ref.[23],
this condition is always satisfied in action (1) with an arbitrary
dilaton and a point electric charge, as long as $B>0$.
Thus the entropy cannot decrease with time
in all the black holes discussed above.

For $N={6\over 5}$, $T_{H}\propto \gamma^{5\over 3}
\Lambda^{{3\over4}}M^{{1\over4}}$ and
$S\propto\Lambda^{-{3\over4}}M^{{3\over4}}$ whereas for $N={4\over 3}$,
$T_{H}\propto\gamma^{3\over 2}\Lambda^{{2\over3}}M^{{1\over3}}$ and
$S\propto\Lambda^{-{2\over3}}M^{{2\over 3}}$. We see that for these black
holes their last breath is cold: {\sl i.e.} $M=0\Rightarrow T_H=0$.
On the other hand, for $N={{6\over 7}}$,
$T_{H}\propto\gamma^{7\over 3}\Lambda^{{3\over 2}}M^{-{1\over 2}}$;
therefore $S\propto\Lambda^{-{3\over2}}M^{{3\over2}}$, the last breath is
hot ($M=0\Rightarrow T_H\to\infty$). Finally, (and perhaps most
interestingly), for $N={4\over5}$, $T_{H}\propto\gamma^{5\over
2}{{\Lambda}^2}M^{-1}$ and $S\propto\Lambda^{-2}M^2$. Thus this
$N={4\over5}$ black hole has the same thermodynamic properties
(apart from the blueshift factor) as the
$(3+1)$-dimensional Schwarzschild one. We believe that this is the first
example of this kind among lower dimensional black holes.

We close this section with some further comments on the temperature. In
ref.[20] it was noted that in ${\cal D}=2n$ ($n\geq 2$) even spacetime
dimensions, Lovelock black holes radiate away with increasing temperature,
while in ${\cal D}=2n-1$ ($n\geq 2$) odd dimensions, they cool off as they
radiate. In particular, the uncharged BTZ black hole (a special case in
ref.[20]) has vanishing temperature as it radiates away all its mass.
In our dilatonic case  with ${\cal D}=3$ we have both situations,
the former for $N>1$ and the latter for $1>N>2/3$. So the additon of a
dilaton indeed changes some of the generic thermodynamic properties of
$(2+1)$-dimensional black holes.

\vskip 0.5 true cm

\section{Cosmological Horizons for $2>N>0$, $A<0$}

Another class of spacetime horizons of physical interest are
the cosmological horizons. It is a well known fact that
for the de Sitter case, a cosmological horizon is present.
However, as we mentioned earlier, when there is a non-trivial
dilaton coupling, the case $\Lambda<0$  no longer behaves as a
de Sitter space. Therefore, it is interesting to see under
what circumstances cosmological horizons may arise.

Similar to section 3 and 4, we still demand that $B>0$, or equivalently
$2>N>0$ and the mass is positive.  The existence of cosmological horizons
is indicated by the fact that as the radial co-ordinate goes large enough
(but still finite), the metric signature flips sign. The limit $r\to\infty$
cannot be taken to define quasilocal mass. However, in the case of
small $\Lambda$ and $M$ [12], we can take the limits
$r^{-{2\over N}+1}\gg A$, $\Lambda r^2\ll 1$ in eq.(24) to identify the
mass parameter in our solution. It is easy to see that the mass reduces to
eq.(25) and so we again identify $A=-{2M\over N}$.

We first consider
the ``massless'' case ({\sl i.e.} $A=0$). Now the metric is
$$
ds^2=-\left({8{\Lambda}r^2\over (3N-2)N}+{8Q^2\over (2-N)N}\right)dt^2
 +{4r^{{4\over N}-2}dr^2\over N^2\gamma^{4\over N}
 \left({8{\Lambda}r^2\over (3N-2)N}+{8Q^2\over (2-N)N}\right)}+r^2d\theta^2.
\eqno(39)
$$
This metric will have a horizon whenever ${\Lambda\over(3N-2)}<0$, with the
(cosomological) horizon located at $r_c=Q\sqrt{{(2-3N)\over (2-N)\Lambda}}$.
When $\Lambda\to 0$, $r_c\to\infty$ and we recover the Shiriashi
solution [10].

We next consider the $\Lambda=0$, $A\neq 0$ case. Now the metric is
$$
ds^2=-\left(-{2M\over N}r^{{2\over N}-1}+{8Q^2\over (2-N)N}\right)dt^2+
    {4r^{{2\over N}-1}dr^2\over {N^2\gamma^{4\over N}(-{2M\over N}r^{{2\over
N}-1}+{8Q^2\over (2-N)N})}}
    +r^2d\theta^2.
\eqno(40)
$$
For positive $M$, there is always a `cosmological' horizon
located at $r_c=({4Q^2\over (2-N)M})^{{N\over 2-N}}$.

Finally, for the neutral case, $Q=0$, and ${2\over 3}>N>0$,
it is easy to show that the cosmological horizon is located at
$r_c=({4\Lambda\over (3N-2)M})^{{N\over 2-3N}}$.
As a matter of fact, we can generally consider all $\Lambda$,
$M$, and $Q$ are non-vanishing. For example, if ${2\over 3}>N>0$
and $\Lambda<0$, a simple graphical analysis shows
that only one real positive root exists for $-g_{tt}=0$
which corrresponds to $r_c$. We will not discuss general
cases in details. If the assumption $B>0$ is released, then further
cosmological or event horizons may exist.
In next section, we briefly discuss
the existence of black holes in such situations.

\vskip 0.5 true cm

\section{Black Holes for $N>2$, $N<0$, $A<0$}

So far we have been assuming $B>0$ in our discussion
on cosmological and event horizons. Physically this means
that the kinetic energy of the dilaton (in flat space) is positive.
In this section, we briefly  point out that
black holes can also arise if the dilaton acts
as a tachyon field ({\sl i.e.} $B<0$, a negative kinetic energy).
Although we can still identify $A=-{2M\over N}$ as the quasilocal mass,
these black holes have a number of rather unattractive and unphysical
properties. Specifically, when the kinetic term is negative, the terms in
the action (1) involving the kinetic, Maxwell and dilaton potential all
diverge as $r\rightarrow\infty$.

It is trivial to see that when $B<0$, one must have $N<0$ or $N>2$ (see
eq.(18)). We will illustrate two cases as examples.

Suppose $N=-2$. Now $g_{tt}=0$ yields
$$
r^4-{2Q^2\over\Lambda}r^2+{2M\over\Lambda}=0.
\eqno(41)
$$
Obviously, the roots are given by
${r_{\pm}}^2={Q^2\over\Lambda}\pm{1\over 2}\sqrt{{4Q^4\over\Lambda^2}
-{8M\over\Lambda}}$ and the extremal condition is given by $Q^4=2M\Lambda$.
One can further calculate the temperature from eq.(36). $T_{H}$ is zero in
the extremal case. When $Q^4<2M\Lambda$, ({\it ie.} for sufficiently large
mass) the horizon disappears. In the pure charge case ($M=0$),
the event horizon is located at $r_h=Q\sqrt{{2\over\Lambda}}$.
The temperature of such a black hole has the property
$T_{H}\propto Q^{3}\Lambda^{-{1\over 2}}$.
As long as $Q$ is a constant, $T_H$ is non-vanishing, and it
keeps radiating. The entropy $S$ is still related to
$r_h$ through eq.(38). Since $B<0$, the area of the event
horizon does not necessary increase with time.
Note that for typical black holes of charge $Q$ and mass $M$,
a naked singularity appears if the charge is too large with respect to
the mass, whereas in this case the situation is reversed.
It seems that the mass is playing the role of
charge and vice versa.

In addition to this, one special black hole solution in the
limit $N\to\infty$ can be derived as follows. When one takes the limit
$N\to\infty$ (or $b\to\sqrt{-2B}$). The dilaton becomes $\phi={2\over
b}ln({r\over\beta})$. Recall that $M$ in eq.(25) is in fact
$M\over\gamma^{2\over N}$. We demand that $\gamma^{2\over N}N$ is finite in
that limit and absorbed in $M$, and the charge $Q$ is large comparable to $N$.
Eq.(20a) now becomes:
$$
ds^2=-(-{2M\over r}+{8\Lambda r^2\over 3}-8{Q}^2)dt^2
     +{4r^{-2}dr^2\over ({-2M\over r}+{8\Lambda r^2\over 3}-8Q^2)}
     +r^2 d\theta^2,
\eqno(42)
$$
where $-N(N-2)$ is absorbed into $8Q^2$ (In fact, it can be checked that
(42) can be alternatively derived if a linear dilaton is assumed, instead
of a logarithm, then perform a co-ordinate transformation to get (42)).  In
this limit, an event horizon exists as long as $\Lambda >0$. For $Q=0$, it
is easily see that $T_H\to\Lambda^{{1\over 3}}M^{{2\over 3}}$. Thus $T_H=0$
when $M=0$.

\section{Conclusions}

We have found a one parameter $(2\geq N>{2\over 3})$ family of static
charged black hole solutions for Einstein gravity minimally coupled to a
dilaton $\phi\propto ln({r\over\beta})$ with an potential
term $e^{b\phi}\Lambda$ for the $2+1$-dimensional action
(1). Their causal structures, Hawking temperature, and entropy were
investigated. One particular black hole ($N={4\over 5}$) has the same
thermodynamic behaviour as the Schwarschild one. Solutions of
cosmological horizons are also found. In the presence of a tachyon field,
black holes are also obtained, which consist of a massless charged black
hole as a particular case. There are no black hole solutions for positive
quasilocal mass if $\Lambda\leq 0$.

One (perhaps unattractive) feature of the solutions (20a)
is that they do not
asymptotically approach anti-de Sitter space. This is because
$\phi$ does not approach a constant in spacelike infinity.
It would be interesting to investigate whether black holes
exist for an asymptotically constant dilaton field.
In this context we note that
the only asymptotically flat black hole solutions
to the Einstein-Hilbert action minimally coupled to
a dilaton with a vanishing potential term, is the Schwarschild
solution ($\phi=0$) [24].
In $2+1$ dimensions, this ``Schwarschild solution''
is not a black hole at all but is instead locally flat spacetime [16].
One should therefore not expect to obtain any asymptotically
flat black hole solution in action (1). The
simplest black hole in (1) is the BTZ one and it is
non-asymptotically flat.

We close by commenting on further possible extensions of our work.
Addition of angular momentum to our solutions is an obvious generalization.
Dimensional reduction of the solutions obtained in this paper to get a new
class of black holes in $1+1$ dimensions would be another interesting
avenue of research.
In the $\phi=0$ case, this was done by Achucarro [25].
On the other hand, one may dimensionally continue
our solutions in the context of Lovelock gravity in
ref.[20]. Regardless, it is always tempting to see how a non-trivial
dilaton field alters the causal structure and thermodynamic
properties of any possible black hole solution.

\section{Acknowledgements}

This work was supported by the Natural Sciences and Engineering Research
Council of Canada. KCK Chan would like to thank Jim Chan for discussions.


\begin{references}
\bibitem[15]{r15}Our factor for the $Q^2ln({\over r_o})$ differs from
	   the one in ref.[1] due to different normalization of the
	   electric charge magnitude.
\end{references}
\end{document}